\documentclass{PoS}

\usepackage{epsfig}

\renewcommand{\d}{\mathrm{d}}

\newcommand{\mev}{~\mathrm{MeV}}

\newcommand{\pp}{\partial}

\newcommand{\lk}{\left<}

\newcommand{\rk}{\right>}

\title{Quark Mass Dependence of the QCD Equation of State on
  $N_\tau=8$ Lattices}

\ShortTitle{Quark Mass Dependence of QCD Equation of State for
  $N_\tau=8$}

\author{\speaker{Wolfgang S\"oldner} {\rm \em for the RBC-Bielefeld
    and hotQCD collaborations}\thanks{This work has been supported in
    part by contract DE-AC02-98CH10886 with the U.S. Department
    of Energy}  \\
  Physics Department, Brookhaven National Laboratory, Upton, NY 11973, USA\\
  E-mail: \email{soeldner@quark.phy.bnl.gov}}

\abstract{We currently perform calculations with an improved staggered
  fermion action (p4fat3). We use a strange quark mass that has been
  tuned to its physical value and light quarks of mass $m_s/20$ on
  lattices of size $32^3 \times 8$. This corresponds to an almost
  physical light quark mass. We present first results on the low
  temperature part of the equation of state of QCD. Through comparison
  with the preliminary hotQCD results on the $N_\tau=8$ equation of
  state, which have been obtained with twice heavier light quark
  masses, we can quantify the quark mass dependence of the equation of
  state in the low temperature regime. We also comment on the quark
  mass dependence of the equation of state at high temperature.}

\FullConference{The XXVI International Symposium on Lattice Field Theory\\
		 July 14-19 2008\\
		 Williamsburg, Virginia, USA}

\begin{document}

\section{Introduction}
\begin{figure}[htp]
  \centerline{ 
    \epsfig{file=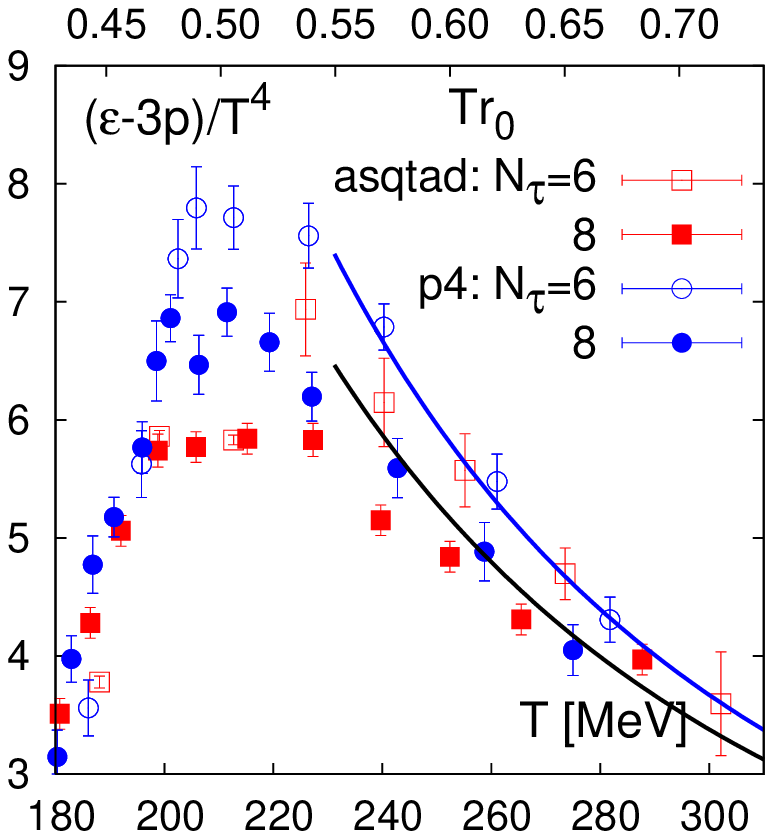, width = 0.5\textwidth}
    \epsfig{file=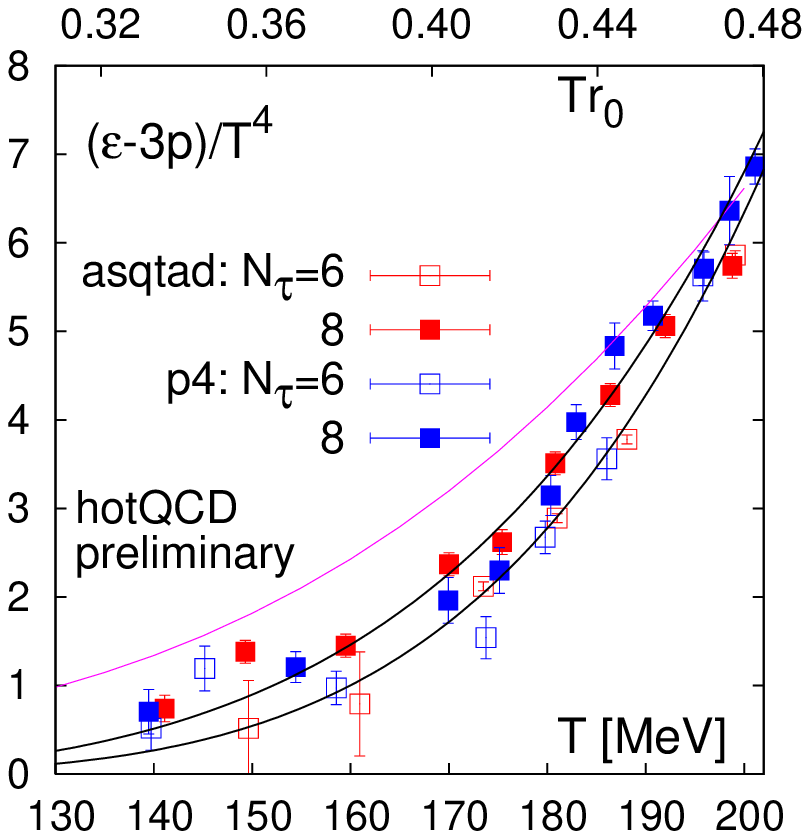, width = 0.5\textwidth}
  }
  \caption{\label{fig:eos}Trace anomaly from lattice QCD at various
    $N_\tau$ for $m_q=0.1m_s$ (left). Deviations are observed between HRG
    and lattice QCD for the low temperature regime (right).}
\end{figure}

One major topic in thermodynamics of QCD and, especially, lattice QCD
is the calculation the Equation of State (EoS). While there has been
great progress in the recent years still several issues remain
unresolved.  One of them is the observed deviation in the EoS of the
lattice computation compared to what one expects from a Hadron
Resonance Gas model (HRG). As can be seen on the right hand side of
Fig.~\ref{fig:eos} the HRG description shows deviations from the
lattice QCD data~\cite{Cheng:2007jq}, especially in the low
temperature regime. On the contrary, because the HRG model is in quite
good agreement with experiment in the low temperature
region~\cite{BraunMunzinger:2003zd}, one would expect that the HRG
model also gives a good description for the low temperature region of
the lattice QCD equation of state.
A possible source for the discrepancy, aside from lattice
discretization errors which still may play a role on the $N_\tau=8$
lattices, may be due to the fact that previous lattice studies of the
EoS had been performed with light quark mass values which are about a
factor two larger than in nature. This corresponds to a light
pseudo-scalar mass of about $220MeV$. In order to quantify the effects
of too large light quark masses we present new calculations for the
lattice QCD equation of state at a smaller light quark mass
$m_q=0.05m_s$ compared to our earlier calculation with $m_q=0.1m_s$
where $m_s$ is the strange quark mass.

The new data for $m_q=0.05m_s$ also is used to investigate the
properties of the chiral phase transition. In particular, we will
present results on the scaling with the light quark mass of the chiral
condensate and the chiral susceptibility which is relevant for
identifying the critical point of the chiral phase transition.

The paper is organized as follows. In Sec.~\ref{sec:details} we give
an overview of numerical details and outline the calculation of the
EoS on the lattice. In Sec.~\ref{sec:pot} the calculation of the
potential and the change in the scale for the $m_q=0.05m_s$ case is
discussed. The results for the EoS with $m_q=0.05m_s$ are presented in
Sec.~\ref{sec:eos} and results for the chiral condensate and
susceptibility are shown in Sec.~\ref{sec:chiral}.

\section{\label{sec:details}Numerical Details and Setup}
The computation of the QCD equation of state on the lattice requires
large resources. Our calculations have been performed on IBM
BlueGene/L and QCDOC supercomputers at Lawrence Livermore National
Laboratory, the New York Center for Computational Sciences (NYCCS), and
J\"ulich Supercomputing Centre. For our calculations we use the p4fat3
action and the RHMC algorithm~\cite{Cheng:2007jq}.  The computational
details are summarized in Tab.~\ref{tab:details} and
Ref.~\cite{Gupta:2008}.  Further information can be found in
Refs.~\cite{DeTar:2008,Karsch:2007,hotQCD:2007}.
\begin{table}[ht]
  \begin{center}
    \begin{tabular}{lll}
      Masses: & $m_q=0.1m_s$ & $m_q=0.00081-0.00370$ \\
      & $m_q=0.05m_s$ & $m_q=0.00120-0.00145$ \\
      \hline
      Pion Masses: & $m_q=0.1m_s$ & $m_\pi \approx220 \mev$ \\
      & $m_q=0.05m_s$ & $m_\pi \approx 160 \mev$ \\
      \hline
      Volume ($T\neq 0$): & $m_q=0.1m_s$ & $32^3 \times 8$ \\
      & $m_q=0.05m_s$ & $32^3 \times 8$ \\
      \hline
      Volume ($T=0$): & $m_q=0.1m_s$ & $32^4$ \\
      & $m_q=0.05m_s$ & $32^4$ \\
      \hline
      Stats. ($T \neq 0$): & $m_q=0.1m_s$ & \# 8,000-37,000\\
      & $m_q=0.05m_s$ & \# 5,000-20,000 \\
      \hline
      Stats. ($T=0$): & $m_q=0.1m_s$ & \#2,000 - 6,000 \\
      & $m_q=0.05m_s$ & \# 1,500-2,000 \\
      \hline
    \end{tabular}
  \end{center}
  \caption{\label{tab:details}Overview of the simulation details.}
\end{table}  

In the following we give a brief sketch of the calculation of the EoS
on the lattice. The basic quantity is the trace anomaly
$\Theta^{\mu\mu} \equiv \varepsilon -3p$ with the energy-momentum
tensor $\Theta^{\mu\nu}$, energy density $\varepsilon$, and pressure
$p$. The trace anomaly can be computed on the lattice in terms of the
action density, and the strange and light quark chiral condensates. We
are only interested in the thermal part of the trace anomaly and
subtract the zero temperature part,
\begin{equation}
  \Theta^{\mu\mu} \equiv \varepsilon -3p = (\varepsilon -3p)_T -
(\varepsilon -3p)_{T=0}.
\label{eq:trace}
\end{equation}
From this expression we can determine the pressure through the
thermodynamic relation $\Theta^{\mu\mu}/T^4 = T \pp (p/T^4) /\pp T$
which is integrated to yield
\begin{equation}
  \frac{p(T)}{T^4} - \frac{p(T_0)}{T_0^4} = \int_{T_0}^T \d \bar T
  \frac{1}{\bar T^5} \Theta^{\mu \mu}(\bar T) \; .
\label{eq:pressure}
\end{equation}
Here $T_0$ is chosen to be in the deep hadronic region where $p(T_0)$
is already exponentially small. The energy density $\varepsilon$ then
is determined given $\Theta^{\mu\mu}$ and $p$. For further details see
Ref.~\cite{Cheng:2007jq}.

\begin{table}[b]
  \begin{center}
    \begin{tabular}{lllll}
      $\beta$ & 3.49 & 3.51 & 3.53 & 3.54 \\
      \hline
      $a_{m_q=0.1m_s}$  \footnotesize{$[fm]$} & 0.1455 & 0.1370 & & 0.1272 \\
      $a_{m_q=0.05m_s}$  \footnotesize{$[fm]$} & 0.1435 & 0.1385 & 0.1306 & \\
      \hline
      $r_0/a$ {\tiny $m_q=0.1m_s$} & 3.223(23) & 3.423(30) &  & 3.689(37)  \\
      $r_0/a$ {\tiny $m_q=0.05m_s$} & 3.2675(90) & 3.386(20) & 3.592(11) &  \\
      \hline
      $T_{m_q=0.1m_s}$ \footnotesize{$[\mev]$} & 169.5(1.2) & 180.0(1.6) & & 194.0(1.9) \\
      $T_{m_q=0.05m_s}$  \footnotesize{$[\mev]$} & 171.85(47) & 178.1(1.1) & 188.90(60) &\\
      \hline
      $\mathrm{\#}_{m_q=0.1m_s}$ & 330 & 140 & & 300 \\
      $\mathrm{\#}_{m_q=0.05m_s}$ & 200 & 200 & 170 & \\
      \hline
    \end{tabular}
    \caption{\label{tab:pot}Results and overview for the calculation
      of the potential. The last two rows give the number of
      configurations we have used in our analysis.}
  \end{center}
\end{table}

\section{\label{sec:pot}Potential}
To account for the quark mass dependence of the lattice cut-off, which
corresponds to a shift in the temperature scale when we lower the
light quark mass, we have calculated the potential for several values
of $\beta$.
\begin{figure}[htp]
  \centerline{
    \epsfig{file=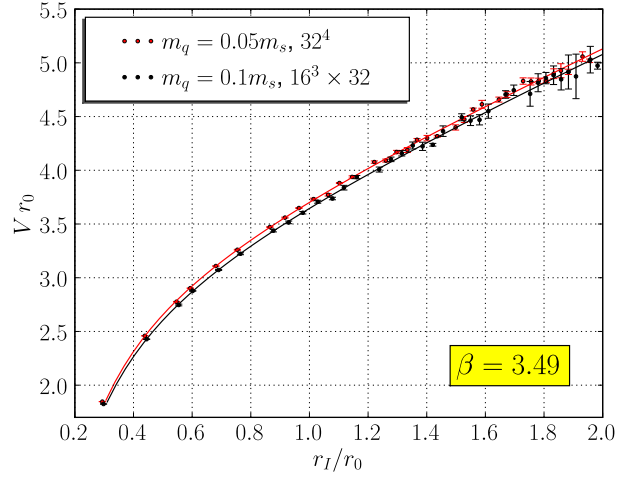, width = 0.6\textwidth}}
  \caption{\label{fig:pot}Potential $V(r)$ vs.~improved distance $r_I$
    at $\beta=3.49$ for $m_q=0.05m_s$ and $m_q=0.1m_s$. Both
    quantities are given in dimensionless units using the Sommer scale
    parameter $r_0$.}
\end{figure}
For $m_q=0.1 m_s$ the scale was calculated on $16^3 \times 32$
lattices, see Ref.~\cite{Cheng:2007jq}. In the case of $m_q=0.05m_s$
we computed the potential on $32^4$ lattices. We have set $r_0 =
0.469$~fm. The results and some simulation details are summarized in
Tab.~\ref{tab:pot} where we also compare lattice scales obtained for
$m_q=0.05m_s$ to $m_q=0.1m_s$. For illustration we plotted the
potential for $\beta=3.49$ in Fig.~\ref{fig:pot} both for
$m_q=0.05m_s$ and $m_q=0.1m_s$.

From the preliminary results in Tab.~\ref{tab:pot} we conclude that
within error bars the scale determined for $m_q=0.05m_s$ is consistent
with the older results for $m_q=0.1m_s$. We are planning to increase
statistics and include more $\beta$ values in the future. For the
moment we will make use of the global mass fit~\cite{Cheng:2006qk}
\begin{equation}
  a_{m_q=0.05m_s} = \exp(-2 A \, \Delta m_q) \, a_{m_q=0.1m_s} \; ,
\label{eq:globalfit}
\end{equation}
in order to determine the scale. The value for $A$ is found to be
$A=1.40(3)$. In this fit more data points are involved than in our
current analysis for $m_q=0.05m_s$. Using this fit we find a shift in
temperature of about $0.5\mev$.

\section{\label{sec:eos}Mass Dependence of the QCD Equation of State}
\begin{figure}[htp]
  \centerline{ \epsfig{file=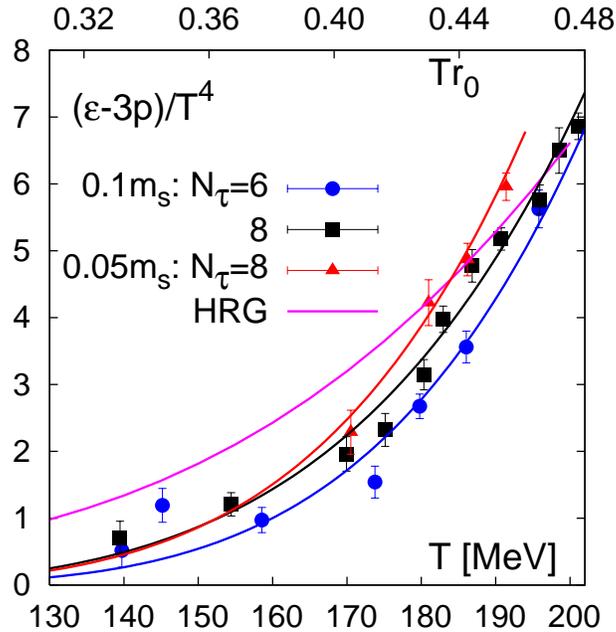, width =
      0.6\textwidth}}
  \caption{\label{fig:eosmass}Trace anomaly for $m_q=0.05m_s$ compared
    to the HRG and $m_q=0.1m_s$ results. The curves are $T^2$ fits to
    the corresponding data made to highlight the cut-off and quark
    mass dependence of the various data sets.}
\end{figure}

In Fig.~\ref{fig:eosmass} we plot our result for the trace anomaly for
$m_q=0.05m_s$. We compare to the $m_q=0.1m_s$ case from
Ref.~\cite{Cheng:2007jq,Gupta:2008} as well as to the HRG model
result. The curves in the plot correspond to fits with a quadratic
ansatz in $T$ which fits the data quite well. Note that for our
calculation of the trace anomaly for the HRG all resonances were
chosen to be at their physical mass. We also note that for the
$m_q=0.05m_s$ case the lightest pseudo-scalar mass\footnote{Note that
  within the staggered formalism only one light pseudo-scalar exists
  at finite values of the lattice cut-off. Full flavor symmetry is
  only recovered in the continuum limit.} is about $160\mev$ which is
very close to its physical value. The shift of about $0.5\mev$ in the
scale, as discussed in Sec.~\ref{sec:pot}, for the $m_q=0.05m_s$ case
has been taken into account.

From Fig.~\ref{fig:eosmass} we observe in the transition region a
shift towards lower temperature for $(e-3p)/T^4$ for $m_q=0.05m_s$ of
about $4\mev$ compared to $m_q=0.1m_s$ (with $N_\tau=8$). For the
larger temperature regime we find that for the smaller quark mass
there is quite good agreement with the HRG model. However, for the
lower temperature region we observe deviations from the HRG in that
case. We will increase the statistics and add more data points in the
lower temperature region in the future in order to clarify the
situation.

\section{\label{sec:chiral}Chiral Condensate}
In this section we present new results on the chiral condensate and
its susceptibility. We consider a combination of the light and strange
quark condensate to eliminate quadratic divergencies with respect to
the additive quark mass renormalization
\begin{equation}
  \label{eq:chiral}
  \Delta_{l,s}=\frac{ \lk \bar\psi\psi
    \rk_{l,\tau} - \frac{\hat m_l}{\hat m_s} \lk \bar\psi\psi
    \rk_{s,\tau}}{ \lk \bar\psi\psi \rk_{l,0} - \frac{\hat
      m_l}{\hat m_s} \lk \bar\psi\psi \rk_{s,0}}.
\end{equation}
We normalize this combination by its zero temperature value to cancel
the multiplicative renormalization factor.
\begin{figure}[htp]
  \centerline{
    \epsfig{file=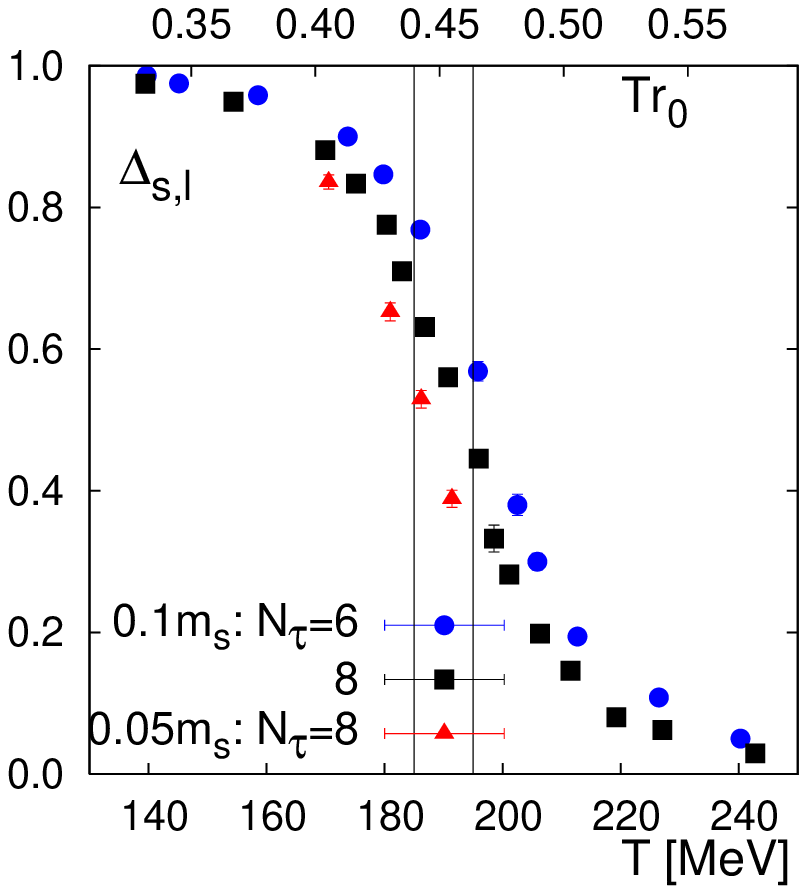, width = 0.5\textwidth}
    \epsfig{file=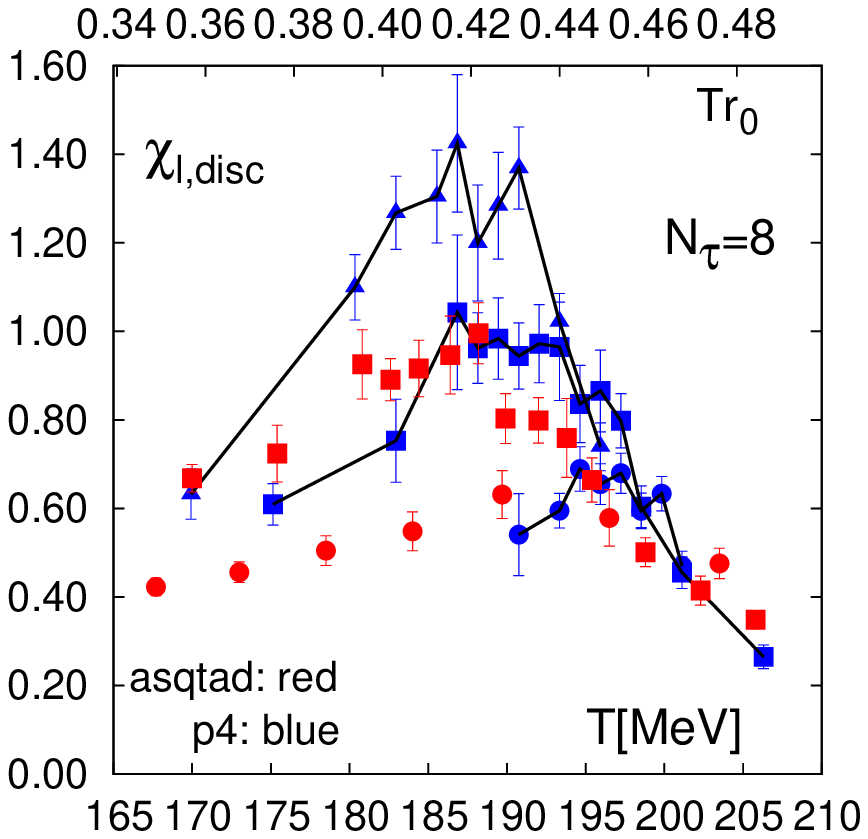,
      width = .5\textwidth}
  }
  \caption{\label{fig:chiralcond} We plot $\Delta_{l,s}$ for the two
    different light quark masses (left). The disconnected light chiral
    susceptibility is shown on the right hand side for light quark
    masses $m_q=0.2 m_s$ (circles), $m_q=0.1 m_s$ (squares), and
    $m_q=0.05 m_s$ (triangles).}
\end{figure}
In Fig.~\ref{fig:chiralcond} we show on the left hand side results for
$m_q = 0.1 m_s$ with $N_\tau=6,8$ as well as results for $m_q = 0.05
m_s$ with $N_\tau=8$. We observe a sharp drop of $\Delta_{l,s}$ in the
region of $T_c$ for both data sets with $m_q = 0.1 m_s$ and $m_q =
0.05 m_s$.  We note that we will present data for more different light
quark masses in the future which will allow us to probe the limit $m_q
\to 0$.  With only two different light quark masses we cannot
investigate this limit at the moment since near $T_c$ terms
proportional to $m_q$ as well as $\sqrt{m_q}$ are present.

On the right hand side of Fig.~\ref{fig:chiralcond} we plotted the
disconnected light chiral susceptibility $\chi_{l,disc}$ for different
light quark masses. We find that $\chi_{l,disc}$ shows a strong quark
mass dependence over a wide temperature range. This can be understood
as an increase in the fluctuations of Goldstone modes below $T_c$. The
transition temperature should then be found near the right edge of the
peak. Further evidence for this picture is given by the observation
that the height of the peak of $\chi_{l,disc}$ scales as $1/\sqrt{m_q}$.

\section*{Summary}
We presented new data for the equation of state for $m_q = 0.05 m_s$
close to the transition temperature. The pion mass is as low as
$\approx 160 \mev$.  We compared the trace anomaly to earlier results
for $m_q = 0.1 m_s$. A total shift of the transition region of about
$4 \mev$ is found towards smaller temperatures. About $0.5 \mev$ of
that shift can be contributed to the shift in the scale. Comparing to
the Hadron Resonance Gas model (HRG) we find that at larger
temperatures the results for $m_q = 0.05 m_s$ are consistent with HRG
model while there are deviations at lower temperatures. Calculations
at different temperatures are currently ongoing.

We have studied the behavior of $\Delta_{l,s}$ for the two different
quark masses. A sharp drop in the vicinity of the critical temperature
is observed. Furthermore, we have presented results for the
disconnected light chiral susceptibility. In the transition region
$\chi_{l,disc}$ shows a broad peak for the different light quark
masses which rises as $1/\sqrt{m_q}$. This gives support for a picture
where the Goldstone modes are causing the fluctuations below $T_c$.

\section*{Acknowledgments}
We are grateful to LLNL, NNSA, New York Center for Computational
Sciences, and the J\"ulich Supercomputing Centre for providing access
to the BlueGene/L supercomputers.

\end{document}